\documentclass{PoS}

\title{A Plan for Electron Ion Collider in China}

\ShortTitle{A Plan for EicC}

\author{\speaker{Xurong Chen}\\
        Institute of Modern Physics, Chinese Academy of Sciences, Lanzhou, 730000, China\\
        E-mail: \email{xchen@impcas.ac.cn}}

\abstract{One of the frontier research areas common to both nuclear and particle physics is the study of hadron structure and the strong interaction. In this contribution we will discuss a plan for a polarized electron-ion collider in China (EicC) and its physics goals.}

\FullConference{XXVI International Workshop on Deep-Inelastic Scattering and Related Subjects (DIS2018)\\
		16-20 April 2018\\
		Kobe, Japan}

\begin{document}


 \section{Introduction}
Electron Ion Colliders (EICs)~\cite{eic}, regarded as a ``super electron microscope", can provide the clearest image inside the nucleon. It is the most ideal tool for us to understand the deep structure of the nuclear matter, especially the quark-gluon structure of the nucleon and nuclei. With high luminosity, broad kinematic range and various polarized ion beams, EIC will be superior to any facility of the past and present time. Polarized EICs are the next generation ``multi-dimensional electron microscopes" that are most effective in studying the deep structure and strong interactions of particles. Currently there are several EIC plans over the world, such as electron Relativistic Heavy Ion Collider (eRHIC)~\cite{erhic}, JLab Electron Ion Collider (JLEIC)~\cite{meic} in USA, and LHeC~\cite{lhec} at CERN (unpolarized). The EIC plans have been discussed in many details in the last decade which are summarized in the INT report~\cite{int} and the EIC whitepaper~\cite{eic}. 
Currently, there are no large-scale accelerator experimental facilities for the study of nucleon structure physics in China. 

\section{EicC Plan and Its Physics Goals}


Based on the HIAF (the Heavy Ion High Intensity Accelerator Facility, approved in 2015 in China), the IMP is proposing to build a high luminosity polarized EIC facility in China, EicC, to carry out the frontier research on nucleon structure studies.

The EicC will be constructed in two stages, i.e. EicC-I and EicC-II, as summarized in Table~\ref{table:parameters}.
\begin{table}[ht]
\centering
\begin{tabular}{|c|c|c|c|c|}
\hline
\hline
Accelerator & Electron beam & Proton beam &$\sqrt{S}$  &Luminosity ($ \mathrm{cm}^{-2} \mathrm{s}^{-1}$) \\
\hline
EicC I & $3 \sim 5$ & $12 \sim 30$ & 12 $\sim 24$ & $4 \times 10^{33}$ \\
\hline
EicC II & $5\sim 10$ & $60 \sim 100$ & 35 $\sim 63$ & $1 \times 10^{35}$ \\
\hline
\hline
\end{tabular}
\caption{EicC-I and EicC-II accelerator design parameters. Both electron and ion beams are polarized. Energy unit in GeV.}
\label{table:parameters}
\end{table}

\begin{figure}[ht]
\centering
  \includegraphics[width=0.6\linewidth]{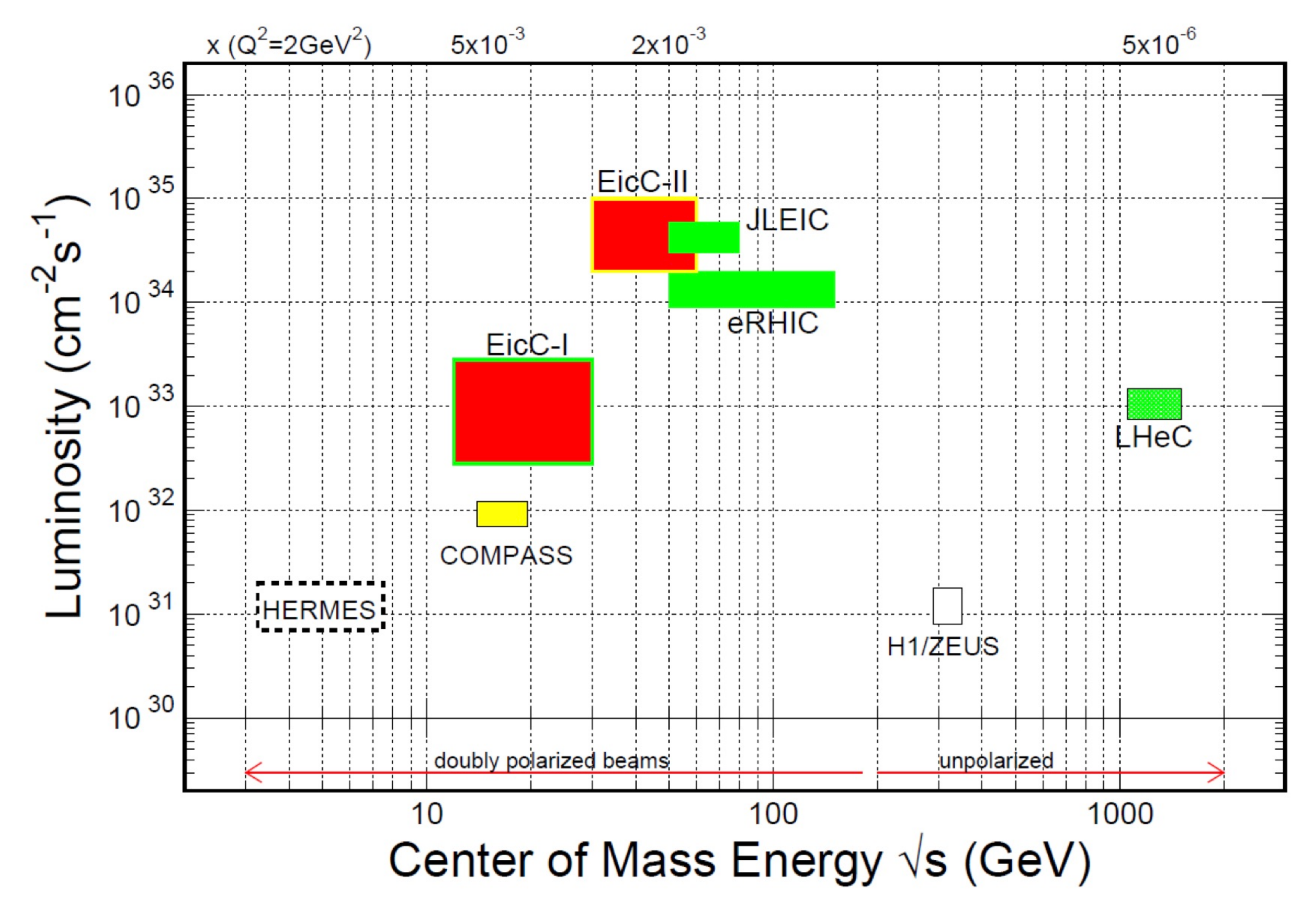}
    \caption{Compare the EicC with non-polarized and polarized electron-nucleon machines. The values of x are also listed in the top of the figure.}
  \label{fig:comparison}
\end{figure}

We compare the EicC with the existing or planned ep facilities in Fig.~\ref{fig:comparison}.

\begin{figure}[ht]
\centering
  \includegraphics[width=0.6\linewidth]{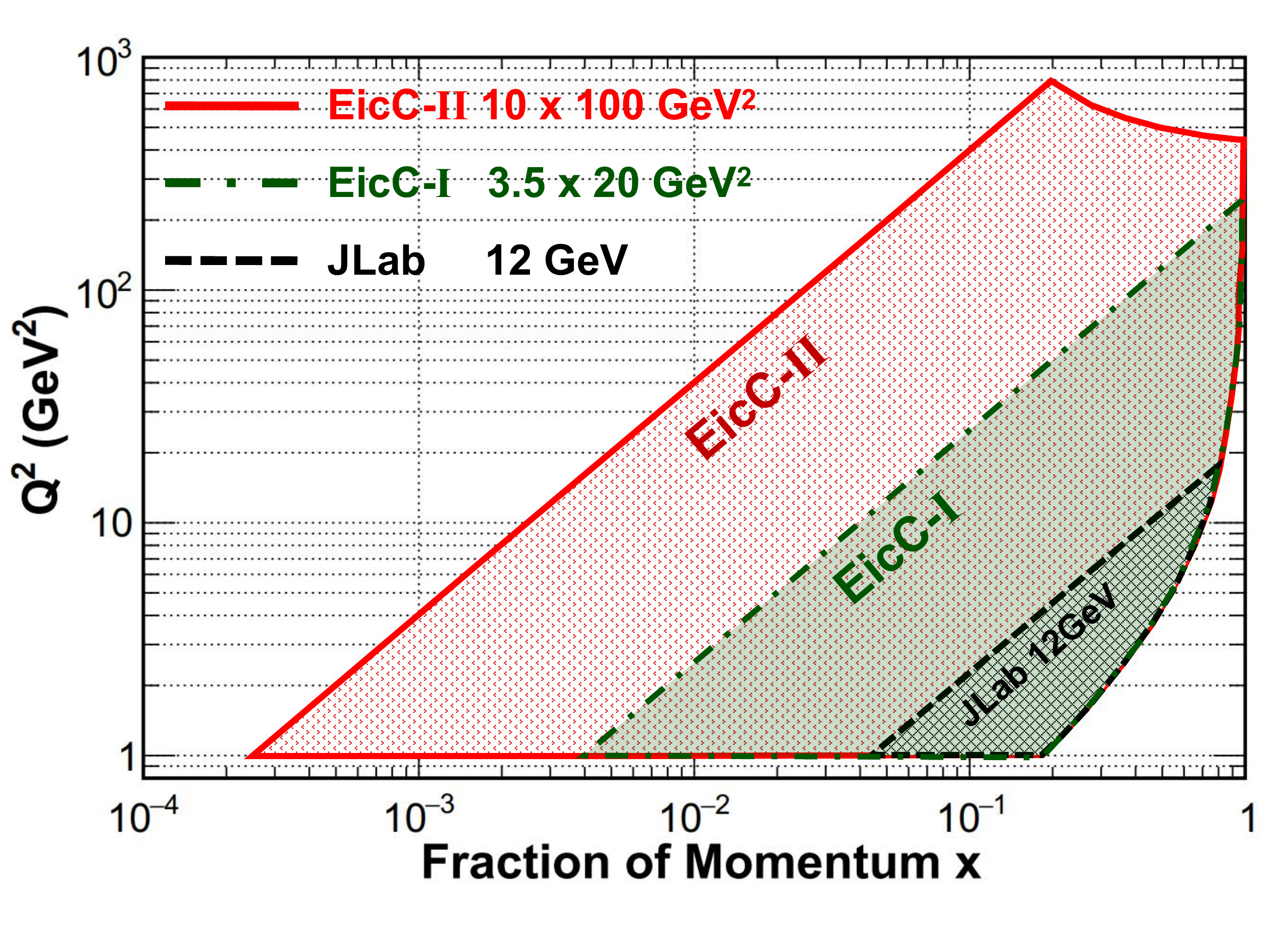}
    \caption{Compare the kinematic ranges of EicC with JLab 12 GeV, for DIS process.}
  \label{fig:kinranges}
\end{figure}

\begin{figure}[ht]
\centering
  \includegraphics[width=0.6\linewidth]{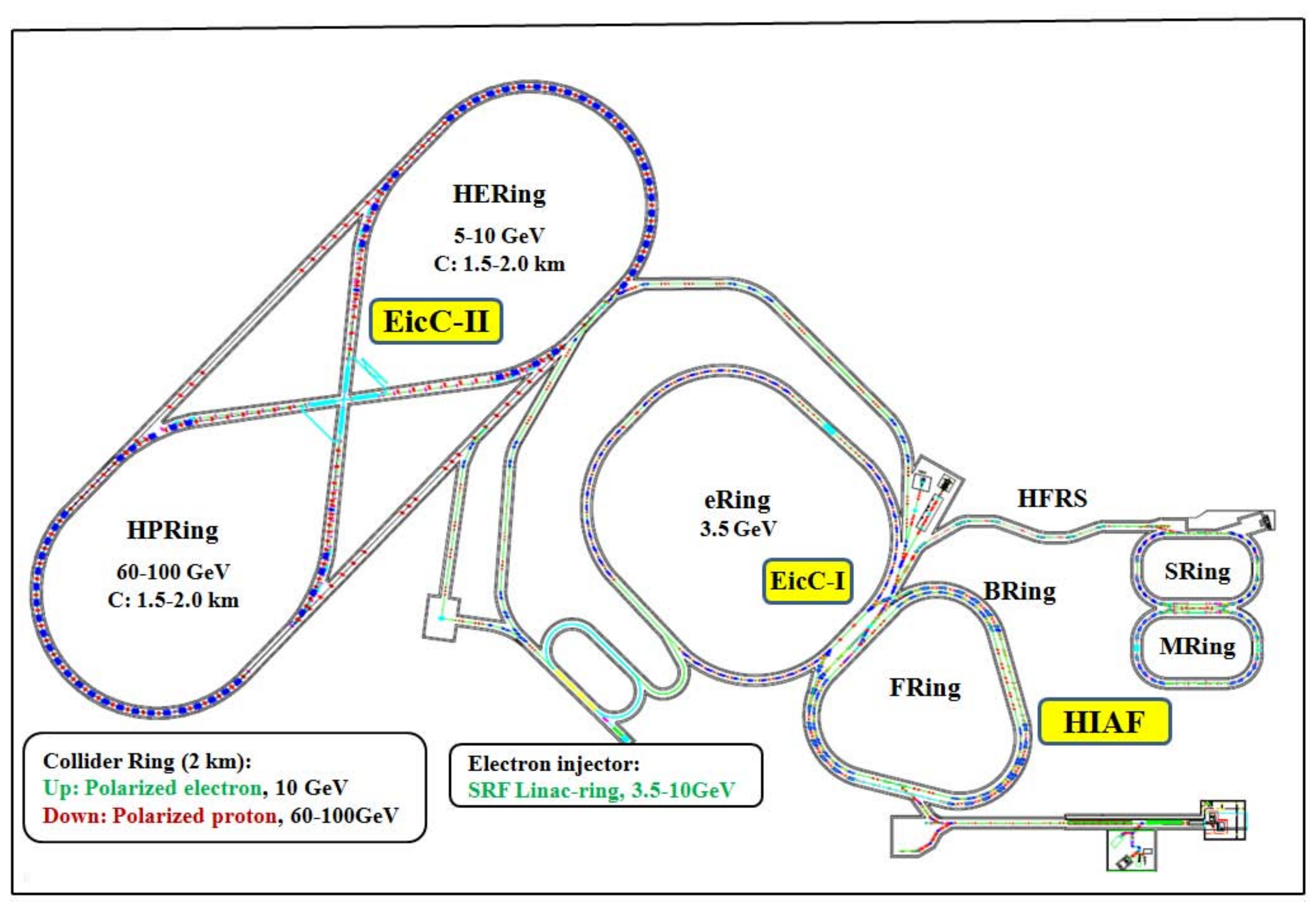}
    \caption{Conceptual layout of EicC-I and EicC-II, based on HIAF.}
  \label{fig:layout}
\end{figure}

In Fig.~\ref{fig:kinranges} we show the kinematic ranges for the EicC and JLab 12 GeV facility. Here we can see that the EicC-I is the best window for sea quark and valence quarks, and the EicC-II can explore gluons and sea quarks structures. The center of mass energy $\sqrt{S}$ is about 16 GeV for the EicC-I. This will provide a clean environment for studying the bottom-hadron, especially the exoctic bottom-quark baryons productions within 2 $\times$ 10 $^{-2}$ $<$ x $<$ 0.5 and 1 $<$ Q$^2$ $<$ 100 GeV$^2$. Unlike other EIC proposals~\cite{eic, erhic, meic, lhec} where the focus is at high energy and small-x in order to probe the gluon's role to the inner structure of nucleons, EicC will focus on valance- and sea-quark contributions to nucleon structure and inter-nucleon effects.

The conceptual layout of EicC-I and EicC-II based on HIAF is presented in Fig.~\ref{fig:layout}. 

\subsection{EicC-I Main Physics Goals}

   The EicC-I covers the range of sea quark, so it will offer the best chance to measure precisely the sea quark 1D and 3D distributions. The EicC-I main physics goals are:
 \begin{enumerate}
 \item High precision quantitative measurements of sea quarks 1D structure
 \item High precision quantitative measurements of sea quark TMD
 \item High precision quantitative measurements of sea quark GPD
 \item Exploring hadronization mechanisms
 \item Systematic studies of baryon states with heavy flavor
 \end{enumerate}
 
\subsubsection{Study of nucleon structure}

\begin{figure}[ht]
\centering
  \includegraphics[width=0.9\linewidth]{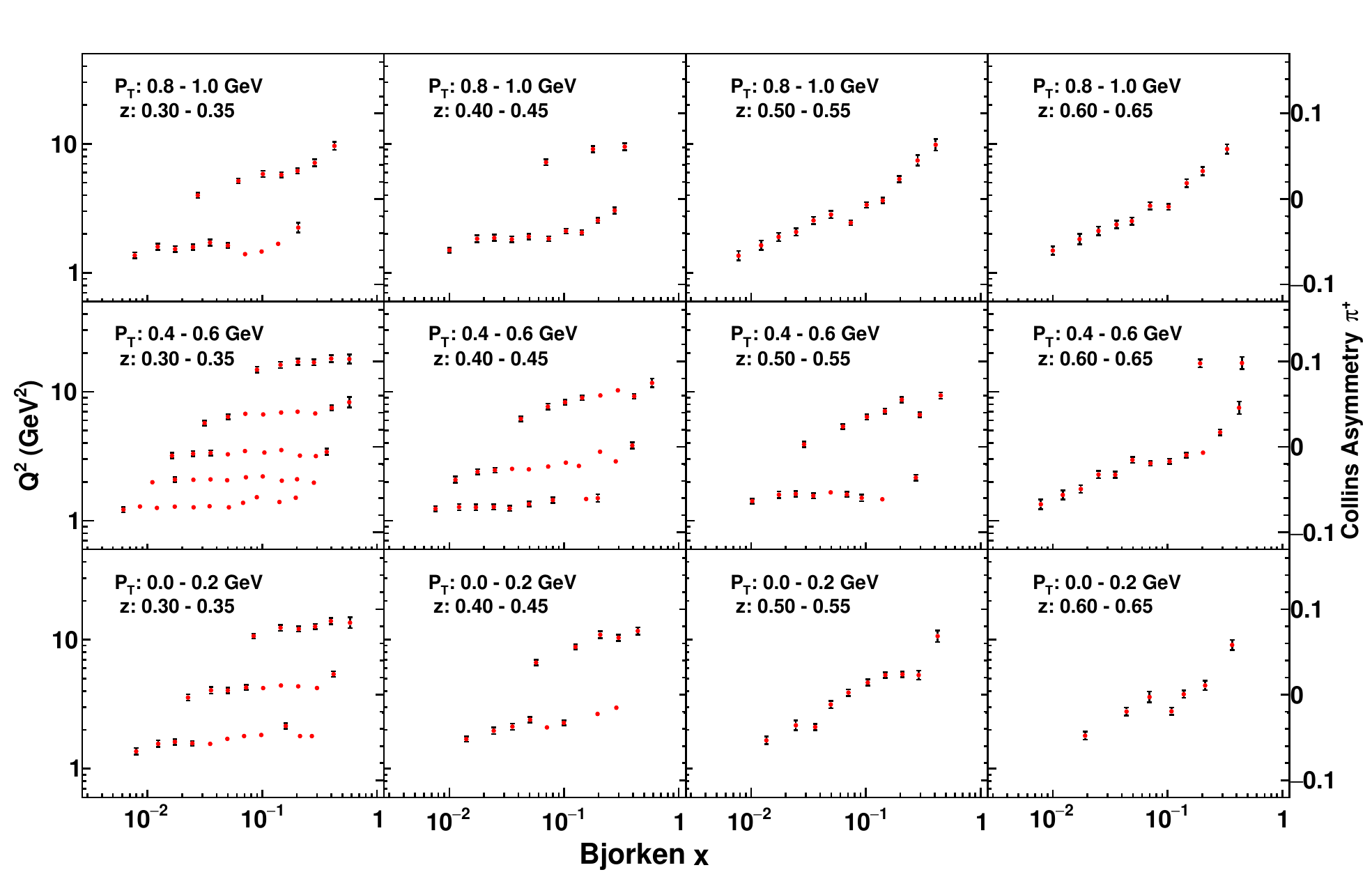}
    \caption{Simulation of TMD Colins asymmetry uncertainty in 4D(x, z, P$_{T}$ and Q$^2$) for one year runing of EicC-I with energy 3.5 $\times$20 GeV.}
  \label{fig:sidis}  
\end{figure}

EicC will perform extensive study on 1D and 3D structure of nucleon and explore the fundamental questions the previous experiments have not answered yet. High-precision measurements of the distribution functions of valence quarks, sea quarks and gluons at EicC will uncover the internal structure of nucleons and ultimately solve the puzzles about nucleon properties. EicC will provide great opportunities to great discoveries in the area of nucleon structure. Fig.~\ref{fig:sidis} shows EicC-I (3.5 $\times$20 GeV) simulation of TMD Colins function asymmetry uncertainty which is pretty small in both valence and sea quark region.

Besides the physics listed above, EicC-I also has the potential to make important contributions to other areas, such as the structure functions of pion and kaon, EMC-SRC effects, etc.

\subsubsection{Systematic Studies of Baryon States with Heavy Flavor}

\begin{figure}[ht]
\centering
  \includegraphics[width=0.7\linewidth]{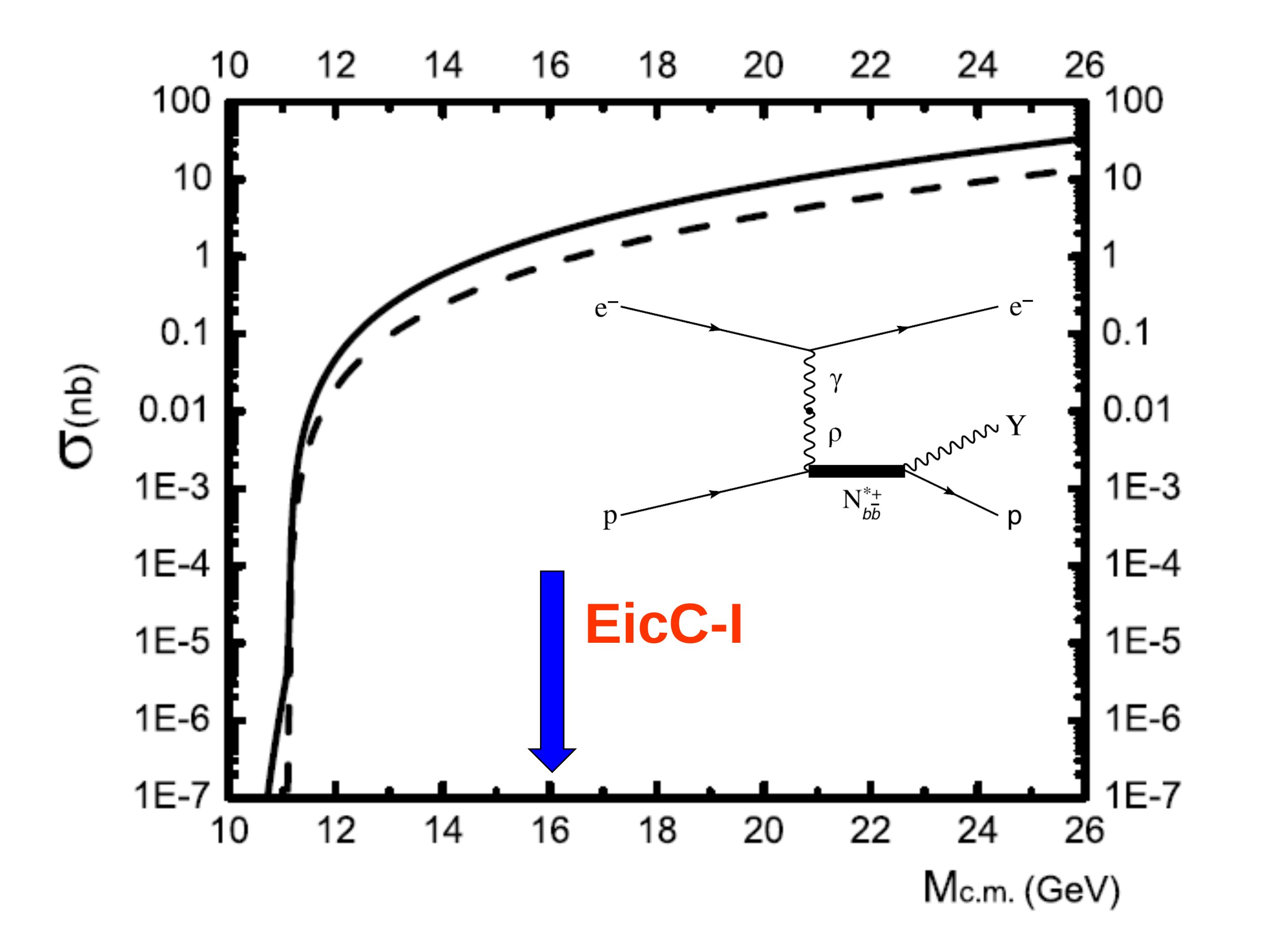}
    \caption{Total cross section of ep $\rightarrow$ eN$^{*+}_{b\bar{b}}$ $\rightarrow$ ep$\Upsilon$~\cite{Zou2012}. The center mass of energy of EicC-I is greater than 16 GeV.}
  \label{fig:bottomonium}
  
\end{figure}

Study of the nucleon and its resonance is an important subject of hadron physics. Tens of exotic states have been observed in the current machines, but their nature is still unclear; some of them are only observed in a specific process. Hence, further studies on the experiment side, are needed. Because of its high energy, high luminosity and low background, especially, the polarized EicC can be used to pin down the quantum number of the observed particles , EicC-I will be an excellent place to study those predicted penta-quark states with 
hidden charm and hidden beauty baryons~\cite{Zou2012,chen2014,chen2016,Meziani1,Meziani2}. Fig.~\ref{fig:bottomonium} shows the total cross section of ep $\rightarrow$ ep$\Upsilon$ via the expected hidden beauty baryon P$_b$ (i.e. N$^{*+}_{b\bar{b}}$) ~\cite{Zou2012}. Meanwhile, EicC-I can also be used to study baryon states with open heavy flavor and other exotic particles.

Another topic is to study non-perturbative gluon. As we know, about 22$\%$ of proton mass comes from trace anomaly, but so far we know very little about it. JLab 12~\cite{jlab12} will measure the J/$\psi$ production near threshold (at small Q$^2$). With high center of mass energy, the EicC will offer precision measurement of $\Upsilon$ production near threshold. the heavier mass of the bottom should help to suppress the theoretical systematic uncertainties. This measurement will shed light on the proton mass study. The third topic is that we can image gluon GPD structures through electro-production of$J/\psi$ and $\Upsilon$ at large value of invariant mass W~\cite{Meziani1,Diehl}.

\subsection{ EicC-II Main Physics Goals}
We plan to build EicC-II based on EicC-I by improving the energy and luminosity. Besides higher-precision measurements of nucleon structures and hadron spectroscopy, there are good prospects for searching for new physics beyond standard model on EicC-II due to its high energy, high luminosity and polarization of electron beam. Physics goals of EicC-II:
\begin{enumerate}
\item Precise measurements of the multi-dimensional spin structure of gluons and sea quarks in the nucleon. 
\item Search for new physics beyond standard model through precise measurements of charged lepton flavor violation processes and the weak mixing angle.
\end{enumerate}

\section{Conclusion}
In summary, EicC will provide an excellent infrastructure for study both sea-quark and gluon contributions to the proton structure. EicC is complementary to JLab 12 GeV and EIC physics programs in U.S. and Europe. The highly polarized EicC plus the advanced high-tech detecting system will allow a series of world-class precision measurements for nucleon structure related physics. In addition to the advancement of the science, the EicC will also greatly boost the technology development for the detector system. In the mean time, a team of young scientists for both physics and technology will also be trained for the future.  

{\bf Acknowledgments:} We are grateful for many enlightening discussions with physics,detector and accelerator working group. Especially thanks to J. P. Chen, N. Xu, Z. Yang and B. S. Zou 
for fruitful discussion and suggestion.

\end{document}